\documentclass[a4paper]{IEEEtran}

\usepackage{footnote}
\usepackage{nopageno}
\usepackage{float}
\usepackage{url}
\usepackage{amssymb,amsmath,amsfonts}
\usepackage{graphicx}
\usepackage{mathptmx}  
\usepackage{textcomp}
\usepackage{xcolor}
\usepackage[absolute,showboxes]{textpos}
\usepackage{multirow}
\usepackage{multicol}
\usepackage{array} 
\usepackage{cite}
\usepackage{algorithm,algorithmic}
\usepackage{amsmath}
\usepackage[cmintegrals]{newtxmath}
\usepackage[utf8]{inputenc}
\usepackage[T5]{fontenc}
\usepackage{tikz}
\usetikzlibrary{positioning}
\usepackage{color,soul}

\begin{document}
\pagenumbering{gobble}
\title{$Q$-MIND: Defeating Stealthy DoS Attacks in SDN with a Machine-learning based Defense Framework}
\author{
	\IEEEauthorblockN{Trung V. Phan\IEEEauthorrefmark{2}, T M Rayhan Gias\IEEEauthorrefmark{2}, Syed Tasnimul Islam\IEEEauthorrefmark{2}, \\ Truong Thu Huong\IEEEauthorrefmark{4}, Nguyen Huu Thanh\IEEEauthorrefmark{4} and Thomas Bauschert\IEEEauthorrefmark{2}}\\
    \IEEEauthorblockA{\IEEEauthorrefmark{2}Chair of Communication Networks, Technische Universit{\"a}t Chemnitz,  09126 Chemnitz, Germany}\\
    \IEEEauthorblockA{\IEEEauthorrefmark{4}School of Electronics \& Telecommunications, Hanoi University of Science and Technology, Hanoi, Vietnam}\\
    \IEEEauthorblockA {Email: trung.phan-van | thomas.bauschert@etit.tu-chemnitz.de}
}

\markboth{ACCEPTED for publication in IEEE GLOBECOM conference 2019}%
{Trung~V.~Phan \MakeLowercase{\textit{et al.}}: $Q$-MIND: Defeating Stealthy DoS Attacks in SDN with a Machine-learning based Defense Framework}

\maketitle

\begin{abstract}
Software Defined Networking (SDN) enables flexible and scalable network control and management. However, it also introduces new vulnerabilities that can be exploited by attackers. In particular, low-rate and slow or stealthy Denial-of-Service (DoS) attacks are recently attracting attention from researchers because of their detection challenges. In this paper, we propose a novel machine learning based defense framework named $Q$-MIND, to effectively detect and mitigate stealthy DoS attacks in SDN-based networks. We first analyze the adversary model of stealthy DoS attacks, the related vulnerabilities in SDN-based networks and the key characteristics of stealthy DoS attacks. Next, we describe and analyze an anomaly detection system that uses a Reinforcement Learning-based approach based on $Q$-Learning in order to maximize its detection performance. Finally we outline the complete $Q$-MIND defense framework that incorporates the optimal policy derived from the $Q$-Learning agent to efficiently defeat stealthy DoS attacks in SDN-based networks. An extensive comparison of the $Q$-MIND framework and currently existing methods shows that significant improvements in attack detection and mitigation performance are obtained by $Q$-MIND.
\end{abstract}

\IEEEpeerreviewmaketitle

\section{Introduction}
\begin{figure}
\centering
\includegraphics[width=0.45\textwidth]{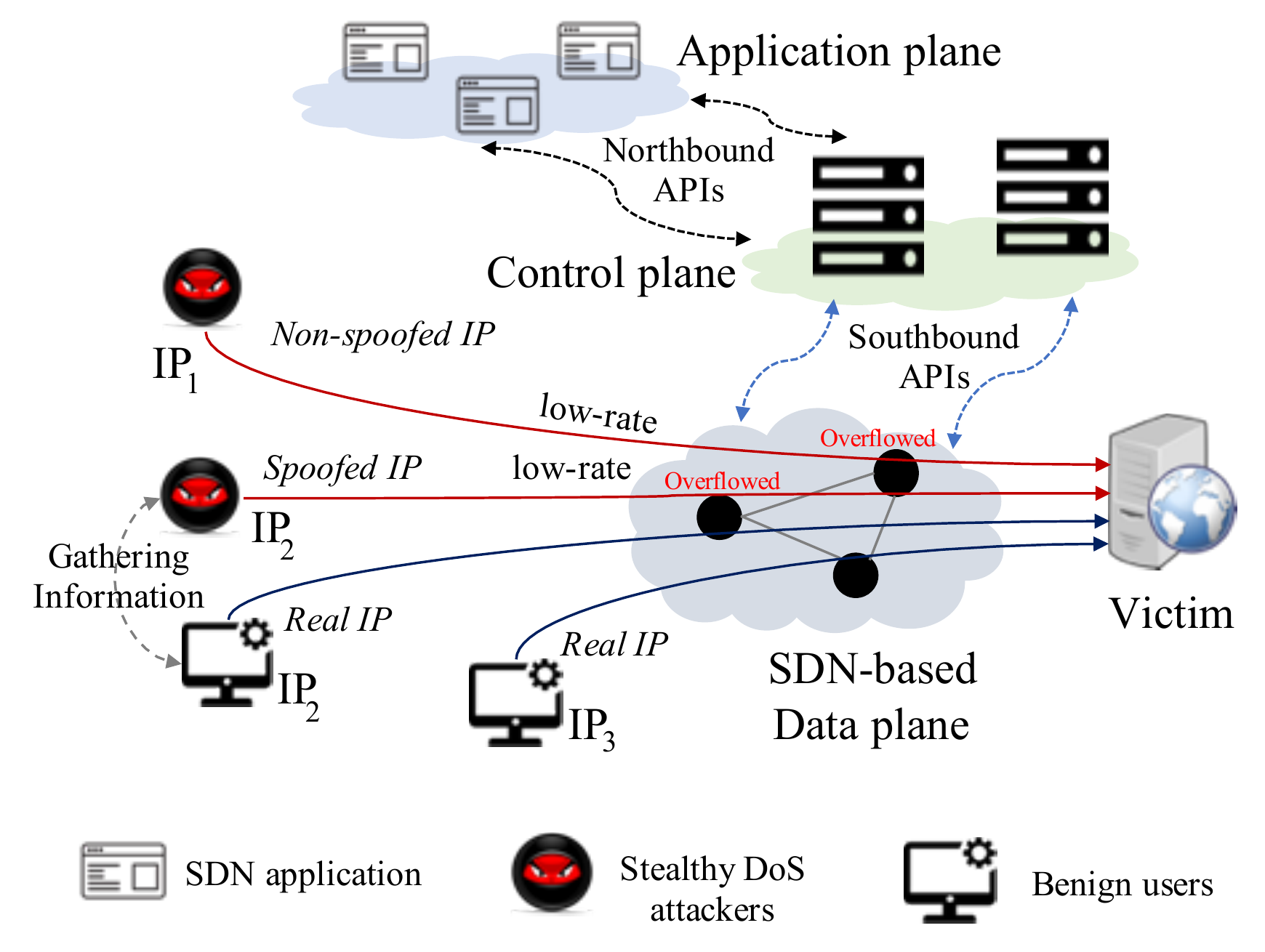}
\caption{Representation of stealthy DoS attacks in the SDN environment}
\label{fig:StealthyDoSinSDN}
\end{figure}

Denial of Service (DoS) and Distributed DoS (DDoS) attacks \cite{DDoSandSDN} can cause serious damage on any networked system. Recently also new DDoS attack variants like stealthy and silent saturation DDoS attacks (also named low-rate and slow DoS attacks) \cite{SlowDoSCategorisation} are observed. Current networks do not yet provide sufficient and efficient countermeasures to defense against these type of attacks \cite{DDoSandSDN}. A global view of the network state and traffic situation is required for effective attack mitigation. The Software Defined Networking (SDN) concept \cite{SurveySDN} is a promising approach to tackle the attack detection and mitigation problem as it can provide a fine grained view on the traffic flows by appropriately setting the flow matching fields in the SDN switches. Although, SDN offers a great potential to defend against novel DoS attack types like stealthy DoS attacks, it is also vulnerable to these attacks, as illustrated in Fig. \ref{fig:StealthyDoSinSDN}. For example, SDN-based forwarding devices, i.e., OpenFlow switches \cite{SurveySDN}, can suffer from overflow problems caused by a silent saturation DoS attack \cite{SlowTCAM,SlowDoSCategorisation,LOFT}.

Stealthy DoS attacks are very difficult to be detected by network operators without accessing the victim machine as the attacker behaves similar to clients with a bad network connection \cite{SlowDDoSMitigation}. Even there are some research efforts \cite{SlowTCAM,LOFT,sdnassistedlowrateddos,http-ddos,FlowtableSharing,ValidateIP} to detect silent DoS attacks, network operations currently still rely on random \cite{SlowTCAM}, predefined threshold-based \cite{http-ddos} or complex high effort \cite{sdnassistedlowrateddos} mechanisms. None of these mechanisms utilize machine learning techniques to provide an early detection of stealthy DoS attacks.

Therefore, in this paper, we propose a novel machine learning based defense framework called $Q$-MIND, to effectively detect and mitigate stealthy DoS attacks in SDN-based networks. We first analyze the adversary model of stealthy DoS attacks, the related vulnerabilities in SDN-based networks and the key characteristics of stealthy DoS attacks. Next, we describe and analyze a detection system that uses a Reinforcement Learning-based approach based on $Q$-Learning in order to maximize its detection performance. Finally we outline the complete $Q$-MIND defense framework that incorporates the optimal policy derived from the $Q$-Learning agent to efficiently defeat stealthy DoS attacks in SDN-based networks.

The paper is structured as follows. Section \ref{StealthyDoSAdversaryModelAnalysis} provides some information about stealthy DoS adversary models and existing countermeasures. Section \ref{SystemModelandProblemFormulation} outlines the architecture of a machine learning based stealthy DoS attack detection system. Section \ref{OptimizingAnomalyDetectionPerformance} presents our approach for optimizing the attack detection performance via $Q$-Learning. The complete $Q$-MIND framework is described in section \ref{Proposal} and the results of the performance analysis are outlined in Section \ref{PerformanceEvaluation}. Section \ref{Conclusion} concludes the paper.

\section{Stealthy DoS Attacks}\label{StealthyDoSAdversaryModelAnalysis}
\subsection{Basic Principle}
Recently a new type of Denial-of-Service (DoS) or Distributed DoS (DDoS) attacks \cite{DDoSandSDN} were identified - stealthy and silent saturation attacks also called low-rate and slow DoS attacks \cite{SlowDoSCategorisation}. These attacks work differently than the high-rate and volumetric DoS attacks. Instead of sending requests to the victim server with a as high as possible rate, attackers periodically send requests with low rate consuming only little resources to render the victim server inaccessible. Stealthy DoS attacks are very hard to be detected in non-SDN networks without accessing the victim machine as the attacker behaves similar to clients with a bad network connection \cite{SlowDDoSMitigation}. During the attack, attackers may exhibit an \textit{ON-OFF} attack pattern which comprises consecutive periods of inactivity (called off-time) and activity (called on-time). Once a stealthy DoS attack has seized all memory space for active connections in a Web server, the attacker tries to keep these connections open as long as possible by exploiting the characteristics of either a specific protocol (e.g., HTTP, DNS) or the application software (e.g., PHP, SOAP) \cite{sdnassistedlowrateddos,http-ddos}.

\subsection{Vulnerabilities in Software Defined Networks w.r.t. Stealthy DoS Attacks}
If the SDN control plane implements a simple flow matching strategy, e.g., only using destination MAC or IP addresses, the ability to track and monitor network traffic for security or forensic analysis is limited. Therefore in order to detect malicious traffic flows stemming from stealthy DoS attacks a more sophisticated flow matching comprising further packet header fields (e.g., IP source addresses) has to be deployed. However, such a flow matching strategy leads to much more flow entries in the SDN switch, so that the maximum number of flow entries might be reached quite soon which then would cause a significant degradation of the forwarding performance or even outage of the switch. Fig. \ref{fig:StealthyDoSinSDN} illustrates stealthy DoS attacks in the SDN environment. An SDN switch is able to simultaneously maintain only a limited number of flow rules, e.g., 2000-3000 flow rules (in case of an OpenvSwitch, \cite{SlowTCAM}) as the flow rules are stored in power-hungry and expensive Ternary Content Addressable Memory. Therefore, attackers can easily compromise an SDN switch by sending new packets which do not match to any current flow rules in the switch. As the current flow rules are preserved, this dramatically increases the number of flow rules in the switch. Consequently, not only the server becomes the victim of a stealthy DoS attack, but also the SDN control and data plane components suffer from resource exhaustion \cite{LOFT,SlowTCAM}. Normally, in case of stealthy DoS attacks the number of flow rules do not exceed the flow-table capacity in the SDN switches. Therefore it is a quite challenging task for an anomaly detection mechanism to detect these type of attacks \cite{LOFT,SlowDDoSMitigation}. To our best knowledge, there are no previous studies that completely solved the stealthy DoS attack detection problem so far.

\subsection{Existing Countermeasures}
The problem of precise detection and mitigation of stealthy DoS attacks is already addressed in the SDN research community. The authors in \cite{SlowTCAM} propose a method which monitors the number of flow entries in SDN switches and, after exceeding a threshold, randomly drops flow rule entries. However there is a probability of dropping flows of legitimate clients as well. The detection technique proposed in \cite{sdnassistedlowrateddos} monitors every incoming flow and calculates suspiciousness scores. But it requires a high effort to store and analyze information of all flows. The authors in \cite{http-ddos} recommend that if the number of incomplete HTTP requests are larger than a threshold, a defense scheme is triggered to drop incomplete request flows. The threshold value might differ from server to server. In addition, in order to prevent attackers from probing idle and hard timeout values in the target SDN-based network, the authors in \cite{LOFT} propose to generate an artificial jitter and set a dynamic timeout for incoming packets. This technique however induces the control plane to process more packet-in messages leading to extra forwarding delays for benign packets. Overflow problems in SDN switches caused by a stealthy DoS attack might be solved by a flow table sharing scheme with neighbor switches \cite{FlowtableSharing} whenever an attack occurs. However, in case of a massive attack, neighbor switches can be flooded by the victim switch and the whole network might suspend operation. In \cite{ValidateIP}, the authors propose to validate source IP addresses by querying the log and if the number of packets/second and the number of bytes/second sent from a source IP address exceed a threshold, the flow rules related to the source IP address are removed in the SDN switch.

All mentioned existing methods for stealthy DoS/DDoS attack detection rely on either random \cite{SlowTCAM}, predefined threshold-based \cite{http-ddos} or high effort \cite{sdnassistedlowrateddos} mechanisms. This motivated us to propose a machine learning based approach to efficiently detect and mitigate stealthy DoS attacks at an early stage. To our best knowledge, our proposal is the first one applying machine learning based detection for stealthy DoS attacks.

\subsection{Characteristics of Stealthy DoS Attacks}\label{MarkedCharacteristics}
One of the key characteristics of stealthy DoS attacks is that it does not matter whether attackers use non-spoofed or spoofed IP addresses to generate malicious requests to a victim server - see Fig. \ref{fig:StealthyDoSinSDN}. This is because in case the SDN network applies a traffic flow matching mechanism comprising layer 3 (IP addresses) and layer 4 (TCP/UDP ports), abnormal source IP addresses have to be used repeatedly to keep flow rules related to attack traffic alive. Otherwise, flow rules of the used sources will be removed after a flow idle$\_$timeout. Therefore, for traffic anomaly detection it is feasible to rely on \textit{source IP addresses} and categorise incoming traffic flows accordingly. This motivated us to develop a \textit{source-based} mechanism for detecting stealthy DoS attacks.

\begin{figure}
\centering
\includegraphics[width=0.49\textwidth]{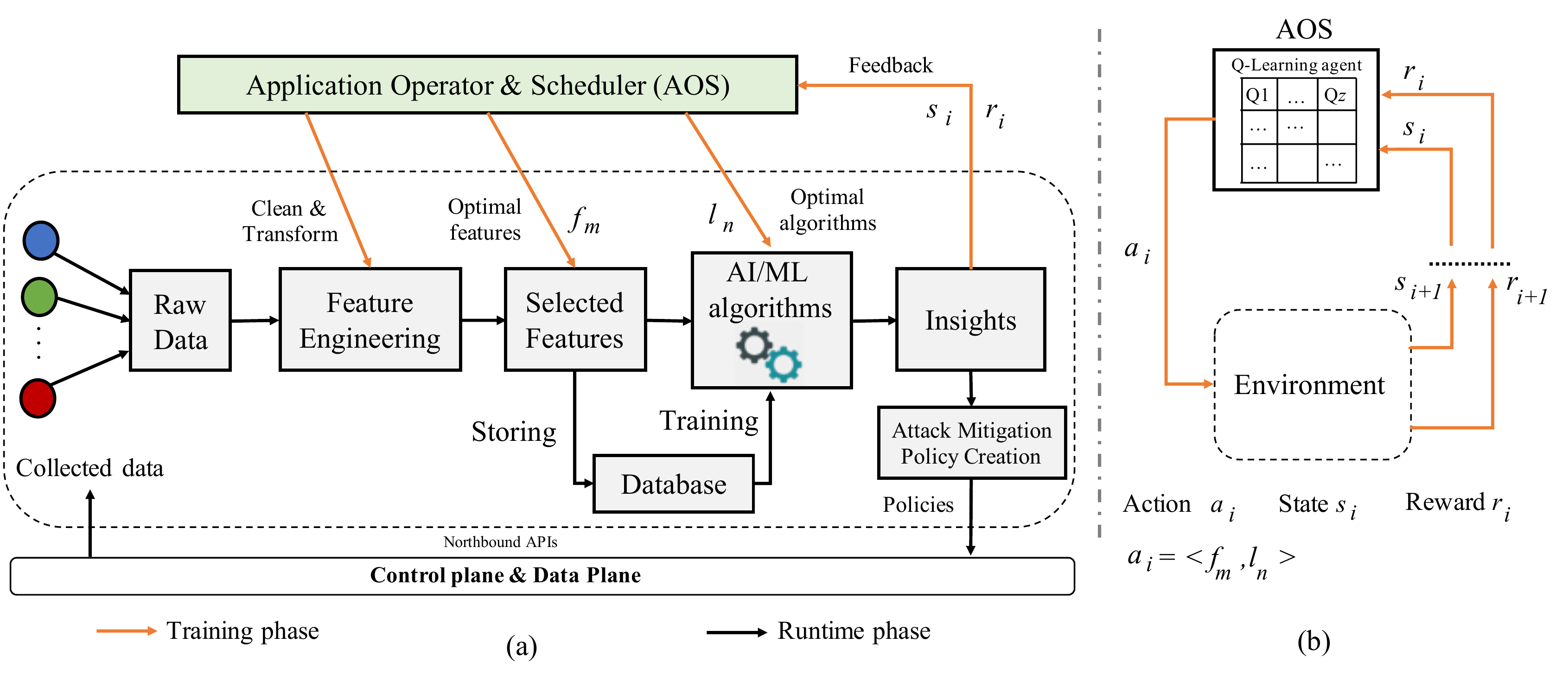}
\caption{Stealthy DoS attack detection system as an SDN application. (a) System overview, and (b) Feature and AI/ML algorithm selection through Reinforcement Learning}
\label{fig:IntelligenceDetectionSystem}
\end{figure}

\section{Basic Architecture of The Stealthy DoS Attack Detection System}\label{SystemModelandProblemFormulation}
Fig. \ref{fig:IntelligenceDetectionSystem} (a) shows the architecture model of a stealthy DoS detection/mitigation system residing in the SDN application plane. Regarding the system operation, first of all data from the network is gathered - for example the northbound APIs (see Fig. \ref{fig:StealthyDoSinSDN}) can be used to query for statistics data from the SDN controller. Detailed statistics information of individual traffic flows in the SDN switches is periodically collected by the SDN controller. Afterwards, the collected data is post-processed in multiple steps. The feature engineering module extracts from the collected data for each source-specific traffic flow a set of features, e.g., the average packets per flow, average packet size per flow, packet change ratio, flow change ratio, etc.. Out of these features the optimum ones (identified by the AOS, see below) are taken and fed into the chosen Artificial Intelligence/Machine Learning (AI/ML) algorithm. The task of the AI/ML algorithm is to classify (based on the selected features) each source IP address into a normal one or malicious one. In case an attack source IP address is recognized the attack mitigation policy creation module formulates a policy for removing malicious source-specific flow rules and blocking the malicious source. The northbound APIs are utilized again to tell the SDN controller to deploy the attack mitigation policies in the data plane.

The application operator and scheduler (AOS) plays an important role in selecting the optimal feature sets and the AI/ML algorithms for an efficient detection operation. The AOS is only active in the training phase, i.e., before the actual runtime operation of the detection system. In the training phase we require a set of labelled traffic data including abnormal and benign samples. This data is either generated by performing simulation experiments of stealthy DoS attacks in an SDN environment or generated from publicly available data sets \cite{CAIDA}. The AOS utilizes the labelled data set to train a chosen AI/ML algorithm with a feature set, and afterwards conducts a cross-validation test to evaluate the attack detection performance for the selected combination of feature set and AI/ML algorithm. Hence, the AOS can adjust these selections to achieve the best attack detection performance. Nonetheless, in practice, selecting an optimal set of features and a suitable AI/ML algorithm is challenging for every classification problem \cite{SelectionApproach1} as there are many possible combinations. Therefore, in the following, we introduce an optimal selection algorithm based on reinforcement learning \cite{ReinforcementLearningIntroduction} that can supervise the optimum selection of the features and AI/ML algorithms.

\section{Optimum Selection of Features and AI/ML-based Classification Algorithms}\label{OptimizingAnomalyDetectionPerformance}
In order to achieve the optimum combination of a certain feature set and a specific AI/ML algorithm w.r.t. the anomaly detection performance, we adopt the $finite$ Markov Decision Process (MDP) approach \cite{ReinforcementLearningIntroduction} with \emph{episodic tasks}. The MDP framework allows the AOS to take an optimal action (combination of feature set and AI/ML algorithm) based on its observations in order to maximize its immediate reward in every single episode. The reward is expressed in terms of multiple evaluation criteria - see below. The MDP is characterized by <$\mathcal{S}$,$\mathcal{A}$,$r$>, where $\mathcal{S}$ is the state space, $\mathcal{A}$ is the action space, and $r$ is the immediate reward of the detection system. For evaluating the anomaly detection performance of an action (feature set and AI/ML algorithm), we consider common metrics \cite{AI-basedTwoStage} including precision ($P_{r}$), recall ($R_{e}$), F-score ($F_{s}$), accuracy ($A_{c}$), and false alarm rate ($F_{a}$). These metrics are calculated from the following observations: TP (True Positive) - number of attacks precisely detected; TN (True Negative) - number of normal patterns precisely classified; FP (False Positive) - number of normal patterns incorrectly classified; and FN (False Negative) - number of attacks unsuccessfully detected. The details of the MDP model are outlined hereafter.

\subsubsection{State Space}
Formally, we can define the state space of the detection system as follows:
\begin{equation}\label{StateSpace}
\mathcal{S} \triangleq \{(P_{r},R_{e},F_{s},A_{c},F_{a})\},
\end{equation}
where $P_{r}=\frac{TP}{TP+FP} \in [0,1]$, $R_{e}=\frac{TP}{TP+FN} \in [0,1]$, $F_{s}=\frac{2}{1/P_{r}+1/R_{e}} \in [0,1]$, $A_{c}=\frac{TP+TN}{TP+TN+FP+FN} \in [0,1]$ and $F_{a}=\frac{FP}{FP+TN} \in [0,1]$.  Then, the state of the detection system is defined as a vector $s$ = $(P_{r},R_{e},F_{s},A_{c},F_{a}) \in \mathcal{S}$.

\subsubsection{Action Space}
$\mathcal{F} = \{f_{1}, f_{2}, ..., f_{m}\}$ denotes a group of feasible feature sets composed of all available and suited features, e.g., a feature set $f_{m}$ consists of 4 features (average packets per flow, average packet size per flow, packet change ratio and flow change ratio). $\mathcal{L} = \{l_{1}, l_{2}, ..., l_{n}\}$ represents a set of possible AI/ML algorithms that can be used for traffic flow classification, e.g., Support Vector Machine \cite{SelectionApproach1}, Random Forest \cite{AI-basedTwoStage}, and Self Organizing Map \cite{SOM}. Then, a tuple, $<f_{m},l_{n}>$, is referred as a combination of a feature set and an AI/ML algorithm. Applying a tuple $<f_{m},l_{n}>$ to the environment, i.e., the detection system - see Fig. \ref{fig:IntelligenceDetectionSystem} (b), means an action is taken by the AOS component. Therefore, the action space is defined as:
\begin{equation}\label{ActionSpace}
\mathcal{A} \triangleq \{a : a = <f_{m},l_{n}>; f_{m} \in \mathcal{F}; l_{n} \in \mathcal{L}\}.
\end{equation}

\subsubsection{Immediate Reward Function}
As aforementioned, we evaluate the anomaly detection performance of a tuple $<f_{m},l_{n}>$ by five criteria. Hence, we define the immediate reward function of the detection system after the AOS takes an action $a$ at state $s$ as the following fitness function:
\begin{equation}\label{RewardFunction}
r(s,a) = W_{P_{r}}P_{r} + W_{R_{e}}R_{e} + W_{F_{s}}F_{s} + W_{A_{c}}A_{c} + W_{F_{a}}e^{-F_{a}},
\end{equation}
where $W_{P_{r}}, W_{R_{e}}, W_{F_{s}}, W_{A_{c}}$ and $W_{F_{a}}$ are weight factors related to the corresponding evaluation criteria, and $W_{P_{r}}+W_{R_{e}}+W_{F_{s}}+W_{A_{c}}+W_{F_{a}}=1$. Note that after performing an action, the AOS observes the feedback from the environment (detection system), i.e., the state vector and the reward value.

\subsubsection{Optimization Formulation}
We define an optimization problem to acquire the optimal policy $\pi^{*}(s)$, being in state $s$ that maximizes the immediate reward in each episode. In particular, in a state $s$ expressed as a vector including $P_{r}$, $R_{e}$, $F_{s}$, $A_{c}$ and $F_{a}$, the policy yields an optimal action or a tuple $<f_{m},l_{n}>$ to maximize the immediate reward of the detection system as defined by Equation \ref{RewardFunction}. The action space $\mathcal{A}$ comprises of $mn$ possible actions. Then, the optimization problem is formulated as follows:
\begin{equation}\label{OptimizationProblem}
\max_{\pi} \{r_{i}(s_{i},\pi (s_{i})): r_{i} \in \mathbb{R}; s_{i} \in \mathcal{S}; \pi (s_{i}) \in \mathcal{A}; 1 \leq i \leq mn\},
\end{equation}
where $r_{i}$($s_{i}$,$\pi (s_{i}$)) is the immediate reward value associated with policy $\pi$ at time step $i$ in an episode.

For solving the optimization problem we apply the $Q$-Learning \cite{ReinforcementLearningIntroduction} algorithm which basically is a Reinforcement Learning approach. By that, the AOS is able to perform an optimal selection without requiring prior knowledge about a set of features and the associated AI/ML algorithm. In other words, we aim to find the optimal policy $\pi^{*} : \mathcal{S} \longrightarrow \mathcal{A}$, i.e., a state-action or state-feature set and AI/ML algorithm mapping table to maximize the anomaly detection performance for stealthy DoS attacks. To achieve these aims, the AOS builds a $Q$-table based on a $Q$-Learning algorithm to store all state-action pair combinations (see Fig. \ref{fig:IntelligenceDetectionSystem} (b)). In a given state $s_{i}$ at iteration $i$ in an episode, the $Q$-Learning agent selects an action $a_{i}$ based on its current selection strategy. Afterwards, it observes the immediate reward $r_{i}$ and the new state $s_{i+1}$, and updates the $Q$-table using a $Q$-function. In other words, the $Q$-Learning agent can learn from its own decisions at each iteration, and it will converge to the optimal policy $\pi^{*}$ after a certain number of iterations \cite{ReinforcementLearningIntroduction}.

Let us denote $\vartheta_{\pi}(s) : \mathcal{S} \longrightarrow \mathbb{R}$ as the expected return of a state $s$ under a policy $\pi$ generally, that is formed as follows:
\begin{equation}\label{ValueFunction}
\begin{split}
\vartheta_{\pi}(s) = \mathbb{E}_{\pi} \left[ \sum_{i=1}^{mn} \gamma r_{i}(s_{i},a_{i}) | s_{i}=s \right] = \mathbb{E}_{\pi} [r_{i}(s_{i},a_{i}) \\ + \gamma \vartheta_{\pi}(s_{i+1}) | s_{i}=s],
\end{split}
\end{equation}
where $\gamma\in$[0, 1) is a discount factor that indicates the importance of the long-term reward \cite{ReinforcementLearningIntroduction}. However, as in our optimization formulation only the immediate reward is considered, $\gamma$ is set to 0 in the remaining of the paper. The optimal policy $\pi^{*}$ in state $s$ represents an action $a$ which yields the maximum value $\vartheta_{*}(s)$:
\begin{equation}\label{OptimalValueFunction}
\vartheta_{*}(s) = \max_{a} \left\{ \mathbb{E}_{\pi} \left[ r_{i}(s_{i},a_{i}) | s_{i}=s \right] \right\}, \forall s \in \mathcal{S}.
\end{equation}
Hence, for all state-action ($s$,$a$) pairs, the optimal $Q$-functions are defined as:
\begin{equation}\label{OptimalQ-Function}
\mathcal{Q}_{*}(s,a) \triangleq r_{i}(s_{i},a_{i}), \forall s \in \mathcal{S}.
\end{equation}
Thus, $\vartheta_{*}(s)$ can be expressed as $\vartheta_{*}(s) = \max_{a} \left\{ \mathcal{Q}_{*}(s,a) \right\}$. By iteratively conducting different actions the optimal value of the $Q$-function \cite{ReinforcementLearningIntroduction}, i.e., $\mathcal{Q}_{*}(s,a)$, for all state-action ($s$,$a$) pairs is found. The $Q$-function is updated at each iteration using the following equation:
\begin{equation}\label{UpdateQ-Function} 
\mathcal{Q}_{i+1}(s_{i},a_{i}) = \mathcal{Q}_{i}(s_{i},a_{i}) + \alpha_{i} [r_{i}(s_{i},a_{i}) - \mathcal{Q}_{i}(s_{i},a_{i})],
\end{equation}
where $s_{i}$ = $(P_{r},R_{e},F_{s},A_{c},F_{a})$, $a_{i}=<f_{m},l_{n}>$ and $\alpha_{i}$ is the learning rate \cite{ReinforcementLearningIntroduction}. The $\alpha_{i}$ value can be a either constant or dynamically adjusted during the learning operation. In addition, to mitigate the exploration and exploitation dilemma that has direct impact on the convergence rate of any learning algorithms, the epsilon-greedy algorithm \cite{ReinforcementLearningIntroduction} is used. Instead of always taking the best action according to the current state, the $Q$-Learning algorithm will then take random actions, and the probability of a random decision is determined by the value of $\epsilon$. Accordingly, the learning operation is terminated when all $Q$ values in the $Q$-table converge.

In conclusion, the $Q$-learning algorithm yields the optimal policy $\pi^{*}(s)$ for a state $s$, i.e., an action $a$, that needs to be taken by the AOS module to maximize the value of the $\mathcal{Q}_{*}(s,a)$ function, i.e., $\pi^{*}(s)=\arg \max_{a}\mathcal{Q}_{*}(s,a)$. Algorithm \ref{OptimalSelectionAlgorithm} provides details of the $Q$-Learning algorithm. 

\begin{algorithm}
\caption{Deriving Optimal Policy with $Q$-Learning}
\label{OptimalSelectionAlgorithm}
\begin{algorithmic}[1]
\STATE \textbf{Inputs}: $\mathcal{F}; \mathcal{L}$; for a state-action pair ($s$,$a$) $\forall s \in \mathcal{S}$, $a \in \mathcal{A}$, then initialize a $Q$-table entry with $\mathcal{Q}(s,a)$ value arbitrarily, $\alpha$ and $\epsilon$, respectively.
\STATE \textbf{begin}
\STATE Repeat the following loop for each episode.
\LOOP 
    \STATE Current state $s_{i}$.
    \STATE Execute action $a_{i}$ according an exploratory policy ($\epsilon$).
    \STATE Obtain the immediate reward $r_{i}$ and new state $s_{i+1}$.
    \STATE Update $\mathcal{Q}$($s_{i}$,$a_{i}$) by using Equation \ref{UpdateQ-Function}.
    \STATE Replace $s_{i} \longleftarrow s_{i+1}$.
\ENDLOOP
\STATE \textbf{Outputs} $\pi^{*}(s) = \arg \max_{a} \mathcal{Q}_{*}(s,a)$.
\end{algorithmic} 
\end{algorithm}

\section{$Q$-MIND Framework}\label{Proposal}
In this section the design and operation of the complete $Q$-MIND framework for detecting and mitigating stealthy DoS attacks is outlined. 
\begin{figure}
\centering
\includegraphics[width=0.47\textwidth]{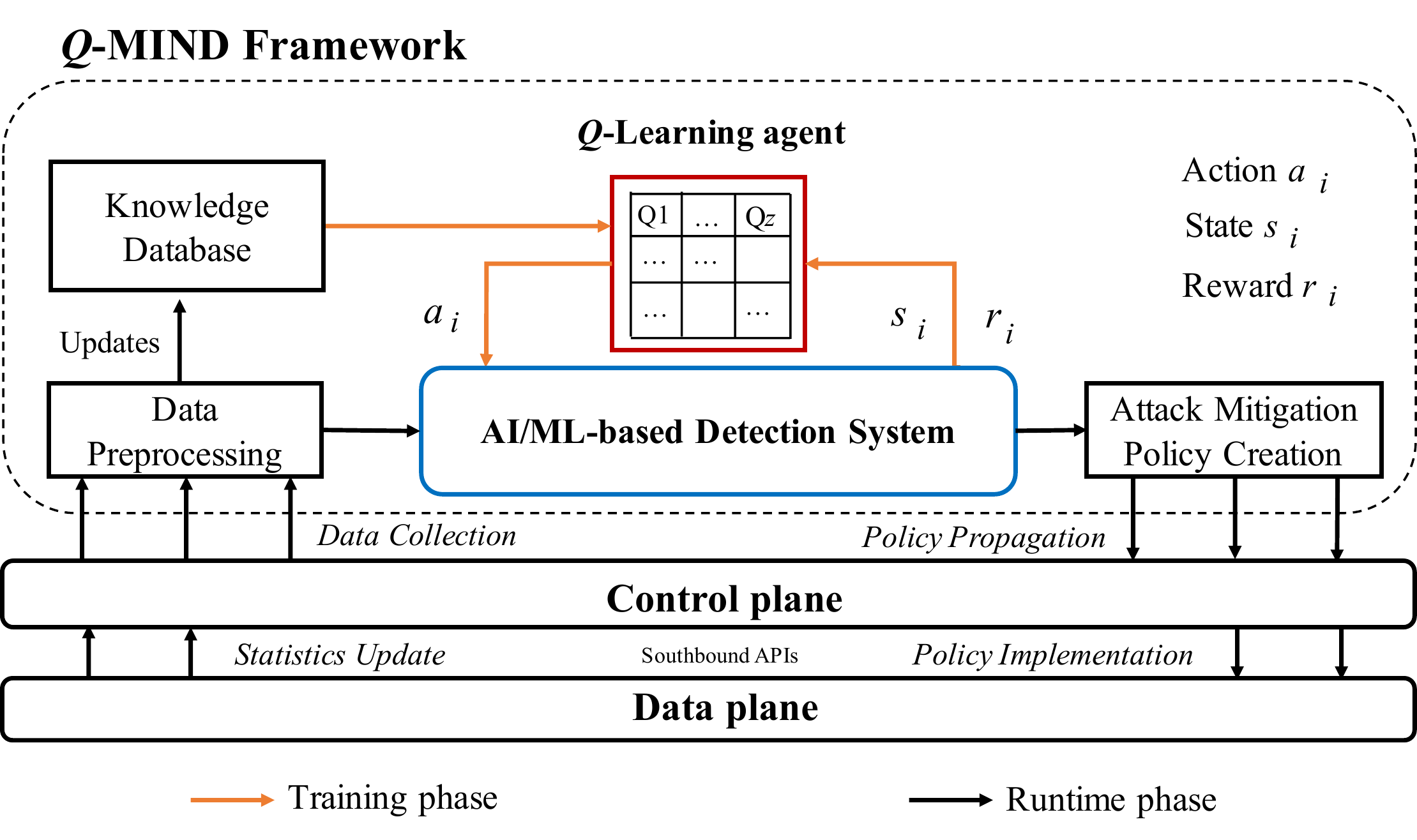}
\caption{$Q$-MIND framework for defeating stealthy DoS attacks in SDN}
\label{fig:ProposedFramework}
\end{figure}

\subsection{$Q$-MIND Architecture}
As depicted in Fig. \ref{fig:ProposedFramework}, the $Q$-MIND framework comprises the following modules: a Data Preprocessing module; a $Q$-Learning agent residing in the AOS component for optimizing the anomaly detection performance; an AI/ML-based anomaly detection system controlled by the $Q$-Learning agent; a Knowledge Database for storing information about feasible features and AI/ML algorithms; and an Attack Mitigation Policy Creation module that issues and implements mitigation rules into the data plane to block stealthy DoS attack traffic.

\subsection{$Q$-MIND Operation}
The operation of $Q$-MIND is described by Algorithm \ref{FightingStealthyDoSinSDN}. First of all, $Q$-MIND runs the $Q$-Learning agent to build a $Q$-table and then generates the optimum action or combination of feature set $f_m$ and AI/ML algorithm $l_n$ as explained earlier. The labeled data set for the training and cross-validation phase is either obtained from simulation experiments of stealthy DoS attacks in an SDN environment or generated from publicly available data sets \cite{CAIDA}. Note that in order to verify the correctness of the anomaly detection, the $Q$-Learning agent conducts cross-validation tests after having trained the detection engine. After the initialization (training and cross-validation) part (lines 1-3) is finished, $Q$-MIND enters the runtime phase detecting and mitigating stealthy DoS attacks by executing the loop part of Algorithm \ref{FightingStealthyDoSinSDN}. The initialization part as well as the loop part can be adjusted lateron in case that further suitable features are identified by the Data Preprocessing module.

\begin{algorithm}
\caption{$Q$-MIND Operation}
\label{FightingStealthyDoSinSDN}
\begin{algorithmic}[1]
\STATE Build a $Q$-table by Algorithm \ref{OptimalSelectionAlgorithm}.
\STATE Derive the optimal action $<f_{m},l_{n}>$ from the $Q$-table.
\STATE Implement a feature set $f_{m}$ and an AI/ML algorithm $l_{n}$ into the detection system.
\LOOP
    \STATE Collect traffic statistics data from the data plane.
    \STATE Extract features from statistics data.
    \STATE Feed features to optimal AI/ML-based detection engine.
    \STATE Get detection result for each source IP address (normal or attack).
    \STATE Create attack mitigation policies if a source IP address is an attacker.
    \STATE Propagate and implement policies to the data plane.
\ENDLOOP
\end{algorithmic} 
\end{algorithm}

\section{Performance Evaluation}\label{PerformanceEvaluation}
In this section, we describe the proof-of-concept implementation of the $Q$-MIND framework and present some results of our performance analysis.
\subsection{Evaluation Scenario Setup}
We perform our experiments by using MaxiNet \cite{MaxiNet} to emulate a simple SDN-based network including a Web server and 8 hosts (4 benign and 4 malicious hosts). The Web server and all hosts are implemented in Linux containers and connect to an OpenFlow switch (OpenvSwitch). The SDN network is controlled by an ONOS SDN controller \cite{ONOS}. We consider three well-known AI/ML based classifiers including Support Vector Machine (SVM-supervised learning) \cite{SelectionApproach1}, Random Forest (RF-supervised learning) \cite{AI-basedTwoStage} and Self Organizing Maps (SOM-unsupervised learning) \cite{SOM}, hence $\mathcal{L}=\left\{SVM,RF,SOM\right\}$. The applied feature set $\mathcal{F}$ is created by the following 10 suitable features: average packets per flow, average packet size per flow, packet change ratio, flow change ratio, average duration per flow, percentage of pair-flows, growth of different ports, average flow inter-arrival-time, fraction of TCP flows over total incoming flows and entropy of incoming flows. These features are extracted for each source IP address taken from a traffic data set including 4000 normal traffic samples and 4000 attack samples. The traffic data set is obtained from our simulation of stealthy DoS attacks in the SDN scenario described above. Accordingly, the $Q$-Learning agent is instructed to train the detection engine and to conduct cross-validation tests. It should be noted that an AI/ML algorithm requires at least 2 features, and that the weight values in Equation \ref{RewardFunction} are set to 0.2 each.

To evaluate the $Q$-MIND framework, we first compare our optimized anomaly detection solution based on $Q$-Learning with three other AI/ML-based anomaly detection/classification methods that apply different feature selection techniques, namely a Principal Component Analysis with a SVM classifier (PCASVM) \cite{SelectionApproach1}, a Generic Algorithm with a SVM classifier (GASVM) \cite{SelectionApproach1} and a Binary Bat Algorithm with a RF classifier (BBARF) \cite{AI-basedTwoStage}. We evaluate the stealthy DoS attack detection performance both in the cross-validation and in the runtime phase. In a second step, we compare the stealthy DoS attack mitigation performance of our $Q$-MIND framework applying the optimal policy and a threshold-based SIFT method \cite{SlowTCAM}. For attack mitigation, we delete all flows stemming from source IP addresses that were identified to belong to attackers and then install flow rules to block these malicious sources for a certain period of time, e.g., 30 seconds.

\subsection{Numerical Results and Analysis}
\subsubsection{Convergence of the selection algorithm (training and cross-validation phase)}

\begin{figure}
\centering
\includegraphics[width=0.43\textwidth]{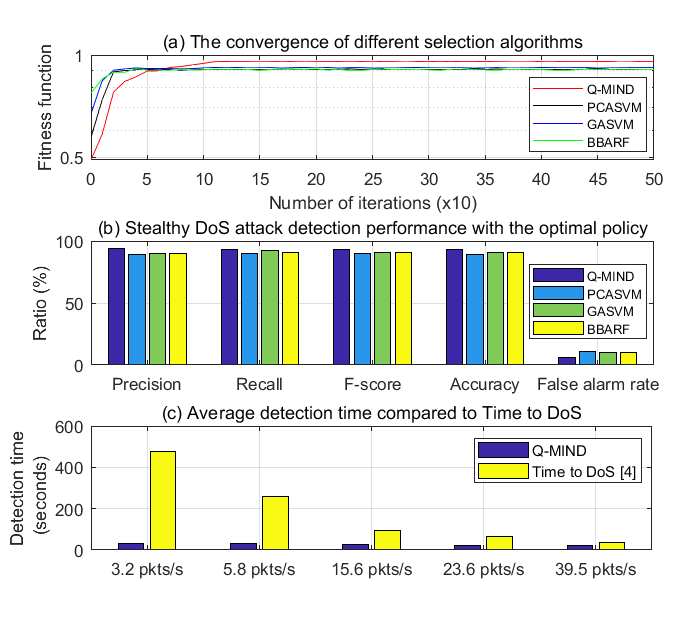}
\caption{Stealthy DoS attack detection performance comparison}
\label{fig:OptimaDetectionPerformance}
\end{figure}

First of all we investigate the stealthy DoS attack detection performance during the training and cross-validation phase of $Q$-MIND. As can be seen in Fig. \ref{fig:OptimaDetectionPerformance} (a), in case of $Q$-MIND, the anomaly detection performance (derived from the fitness function in Equation \ref{RewardFunction}) fluctuates during the first 100 iterations of the training and cross-validation phase because the $Q$-Learning agent frequently updates its $Q$-table in the beginning of the learning phase. Thereafter, it becomes stable and achieves a value of 0.955 for the optimal policy. The other anomaly detection schemes (that operate with fixed classification algorithms and different feature selection techniques) perform well in the first iterations but do not improve anymore in the remaining time. As $Q$-MIND is able to apply different combinations of classification algorithms and feature sets due to the $Q$-Learning approach, it finally finds the optimum policy that yields the best detection performance. In the considered scenario the optimal action turns out to be a combination of a 4-feature set and the $SOM$.

\subsubsection{Attack detection performance applying the optimal policy (runtime phase)}

In the next step we perform experiments in the MaxiNet emulation framework to evaluate the stealthy DoS attack detection performance of $Q$-MIND in the runtime phase. The results are shown in Fig. \ref{fig:OptimaDetectionPerformance} (b). It can be observed that using the optimal policy, the $Q$-MIND framework outperforms the three other methods in all five evaluation criteria and achieves results which are near to the optimal results obtained in the cross-validation phase with the optimal policy. In \cite{SlowTCAM}, it is reported that a stealthy DoS attack where 39.5 unique packets/s are sent to an SDN switch (OpenvSwitch) causes the switch to be overflowed (table-full event) after just 38.0 seconds (Time to DoS). We record the average detection time of the $Q$-MIND framework and compare it with the Time to DoS values reported in \cite{SlowTCAM}, as shown in Fig. \ref{fig:OptimaDetectionPerformance} (c). The results show that the $Q$-MIND framework takes much less time to detect stealthy DoS attacks with different attack rates and that it completely avoids the overflow problem in the switch.

\begin{figure}
\centering
\includegraphics[width=0.43\textwidth]{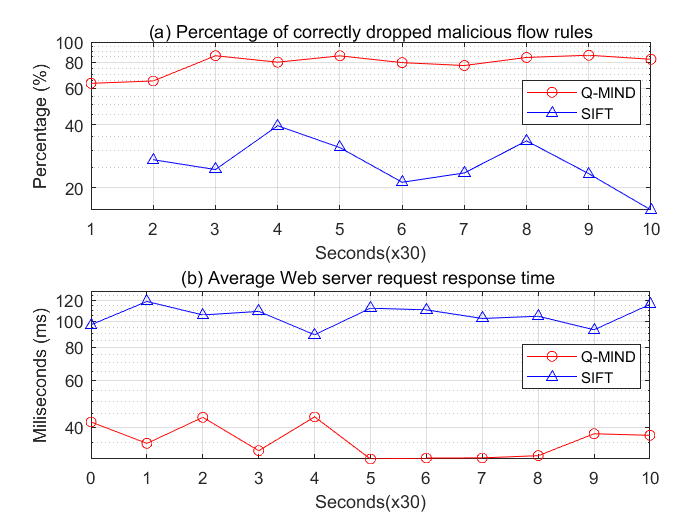}
\caption{Stealthy DoS attack mitigation performance comparison}
\label{fig:MitigationPerformance}
\end{figure}

\subsubsection{Attack mitigation performance}
In order to evaluate the attack mitigation performance we measure the percentage of correctly dropped malicious flow rules in the switch and the request response time of the Web server when the network is under attack. As shown in Fig. \ref{fig:MitigationPerformance} (a), $Q$-MIND achieves a very good percentage of correctly dropped attack flows because it implements policies as soon as a source IP address is detected to stem from an attacker. Contrary the SIFT method randomly drops flows only after the switch gets overflowed, hence malicious flows always remain in the switch. From Fig. \ref{fig:MitigationPerformance} (b) one can see that $Q$-MIND also guarantees an acceptable response time and that it again outperforms the SIFT method.

\section{Conclusion}\label{Conclusion}
In this paper, we propose a novel machine learning based framework, named $Q$-MIND, to effectively defense against stealthy DoS attacks in SDN-based networks. We conduct a comprehensive analysis of the anomaly detection system which incorporates a Reinforcement Learning scheme ($Q$-Learning algorithm) to maximize the anomaly detection performance. Our performance evaluation results demonstrate that $Q$-MIND applying the optimal policy from the $Q$-Learning agent achieves a higher stealthy DoS attack detection and mitigation performance than currently existing methods.

\section{Acknowledgment}
This work has been performed in the framework of the Celtic-Plus project SENDATE Secure-DCI, funded by the German BMBF (ID 16KIS0481).

\bibliographystyle{ieeetr}
\bibliography{references.bib}
\end{document}